\documentclass[twocolumn,prl,preprintnumbers,superscriptaddress]{revtex4-1}
\usepackage{revsymb,graphicx,amssymb,amsmath,natbib,placeins,bm,color,float} 
\usepackage{bbold,bbm,color,rotating,verbatim,multirow,comment,blkarray, 
leftidx}
\usepackage[english]{babel}
\allowdisplaybreaks

\begin{document}
\title{Cascading Failures as Continuous Phase-Space Transitions}

\author{Yang Yang}
\affiliation{Department of Physics and Astronomy, Northwestern University, Evanston, IL 60208, USA}

\author{Adilson E. Motter}
\affiliation{Department of Physics and Astronomy, Northwestern University, Evanston, IL 60208, USA}
\affiliation{Northwestern Institute on Complex Systems, Northwestern University, Evanston, IL 60208, USA} 

\begin{abstract}
\noindent
In network systems, a local perturbation can amplify as it propagates, potentially leading to a large-scale cascading failure. Here we derive a continuous model to advance our understanding of cascading failures in power-grid networks.  The model accounts for both the failure of transmission lines {\it and} the  desynchronization of power generators, and incorporates the  transient dynamics between successive steps of the cascade. In this framework, we show that a cascade event is a phase-space transition from an equilibrium state with high energy to an equilibrium state with lower energy, which can be suitably described in closed form using a global Hamiltonian-like  function. From this function we show that a perturbed system cannot always reach the equilibrium state predicted by quasi-steady-state cascade models,  which would correspond to a reduced number of failures, and may instead undergo a larger cascade. We also show that in the presence of two or more perturbations, the outcome depends strongly on the order and timing of the individual perturbations. These results offer new insights into the current understanding of cascading dynamics, with potential implications for control interventions.
\end{abstract}

\maketitle 

Cascading processes underlie  a myriad of network phenomena~\cite{review},  including blackouts in power systems~\cite{2003report1,2003report2},  
secondary extinctions in ecosystems~\cite{Dunne2009,sahasrabudhe2011rescuing},  and 
complex contagion in financial networks~\cite{gai2010contagion,elliott2014}. 
In all such cases, an otherwise small perturbation 
may propagate  and eventually cause a sizable portion
of the system to fail. Various system-independent cascade models have been 
proposed~\cite{watts2002simple,motter2002,goh2003sandpile,Crucitti2004,buldyrev2010catastrophic,Brummitt2015} and used to 
draw general conclusions, such as on the impact of interdependencies~\cite{brummitt2012suppressing}  and countermeasures~\cite{motter2004cascade}. 
There are outstanding questions, however, for which it is necessary to model the cascade dynamics starting from the actual dynamical state of the system.

In power-grid networks, the state of the system is determined by the power flow 
over transmission lines and the frequency of the power generators, 
which must be respectively below capacity and synchronized under normal steady-state conditions. 
Although a local perturbation has a limited  impact on the connectivity of the network, 
it may 
trigger a cascade of failures  and protective responses  that switch off  grid components  and may also lead generators to lose synchrony.
Much of our current understanding about this process
has been derived from quasi-steady-state cascade models~\cite{dobson2007complex,anghel2007stochastic,conf_rev,matpower,witthaut2015critical,moussaw17},
which use iterative procedures to model the successive inactivation of network components caused by power flow redistributions, while 
omitting  the transient dynamics between steady states as well as the dynamics of the
generators. Further understanding has resulted from stability studies focused on the synchronization dynamics of power generators
in the absence of flow 
redistributions~\cite{susuki2011coherent,rohden2012self,motter2013spontaneous,dorfler2013synchronization,menck2014dead}.

Yet, to date no theoretical approach has been developed  
to incorporate at the same time these two fundamental aspects of power-grid 
dynamics---frequency change and flow redistribution---in the modeling of cascading
 failures~\cite{data}.
The goal of our study is to fill this gap
and consider the interaction between these two factors.
Our framework is inspired by  energy function analysis approaches 
considered in the study of power system stability~\cite{pai2012energy,bergen1981structure} and 
of bistability of circuit elements~\cite{demarco2001phase}. 

Specifically, in this Letter we introduce a time-continuous cascade model
that includes the dynamics of the state 
variables---governed by the swing equations  of the generators, frequency dependence of loads, and power flow equations---as well as
the dynamics of the status variables describing the on/off (i.e., operational/disabled) condition of the transmission lines. 
Within this model,  the steady operating states of
the system
correspond to stable equilibria,  
and a cascade event is a phase-space transition from one stable equilibrium to another. 
We study  these states  and show  
that the stable equilibria are the local 
minima of an energylike function.     
This leads to numerous important implications that have not been systematically studied before. In particular, it 
follows from the properties of this function that a
perturbed system cannot always reach the equilibrium state predicted by quasi-steady-state
 models,
and may instead approach an equilibrium
corresponding to a larger cascade; 
 this highlights the importance of the dynamics between successive steps of a cascade, as considered
in our continuous model, which is a factor that has remained unexplored with few exceptions~\cite{review,helbing2008,cornelius2013,timme2017}.
It also follows  that the equilibrium energy does not depend monotonically on the number of failures,
and that cascades triggered by multiple perturbations depend strongly on the perturbation order.
These results suggest the possibility of cascade  
mitigation using judiciously designed
perturbations to steer the system to a preferred  equilibrium that would not be reached spontaneously.

We first consider the protective operation, 
common to most power networks, 
that {\it  removes} a transmission line when the flow on it exceeds its capacity. 
We associate each line $\ell$ with
a continuous variable $\eta_\ell$
representing its on/off status (as well as the continuous
process of switching between the two conditions) and a parameter $\lambda_\ell$ 
indicating the fraction of the line capacity used by the flow. 
As shown below, this allows 
 us to incorporate the
line status  into the dynamical equations by scaling the power flow terms
by $\eta_\ell$, with $\eta_\ell$ representing the normal status for $\lambda_\ell<1$ and the failed status for $\lambda_\ell\ge1$, where
$\eta_\ell$  is thus constrained to the unit interval. 
To model  the automatic removal of the overloaded lines, we 
can then define the dynamics of $\eta_\ell$~as
\begin{equation} \label{eq:eq1}
\dot{\eta}_\ell
 =  f(\eta_\ell) - \lambda_\ell,
\end{equation}
where  the 
rhs~is defined to satisfy three physical conditions: 
(I)~for $\lambda_\ell<1$, there are three equilibria $\eta_\ell^{\text{(f)}}<\eta_\ell^{\text{(c)}}<\eta_\ell^{\text{(n)}}$, 
where $\eta_\ell^{\text{(n)}}\approx1$ is a stable equilibrium  representing the normal operation 
status,  $\eta_\ell^{\text{(f)}}\approx0$ is a stable equilibrium  representing the failed status, and
$\eta_\ell^{\text{(c)}}$ is an unstable equilibrium marking the critical value below which $\eta_\ell$ evolves to the failed status;
(II)~for $\lambda_\ell\ge1$, only the equilibrium $\eta_\ell^{\text{(f)}}$ remains stable, which is satisfied if the local maximum of $f$ in $(\eta_\ell^{\text{(c)}},\eta_\ell^{\text{(n)}})$ is $1$;
(III)~$\eta_\ell^{\text{(c)}}$ is always close to $1$, since a line should be fully operational under normal conditions.
The dynamics does not depend sensitively on the details of function $f$ provided these conditions are satisfied.
 Throughout, we use overdot to indicate time derivative.

\begin{figure}[t]  
\includegraphics[width=8.3cm]{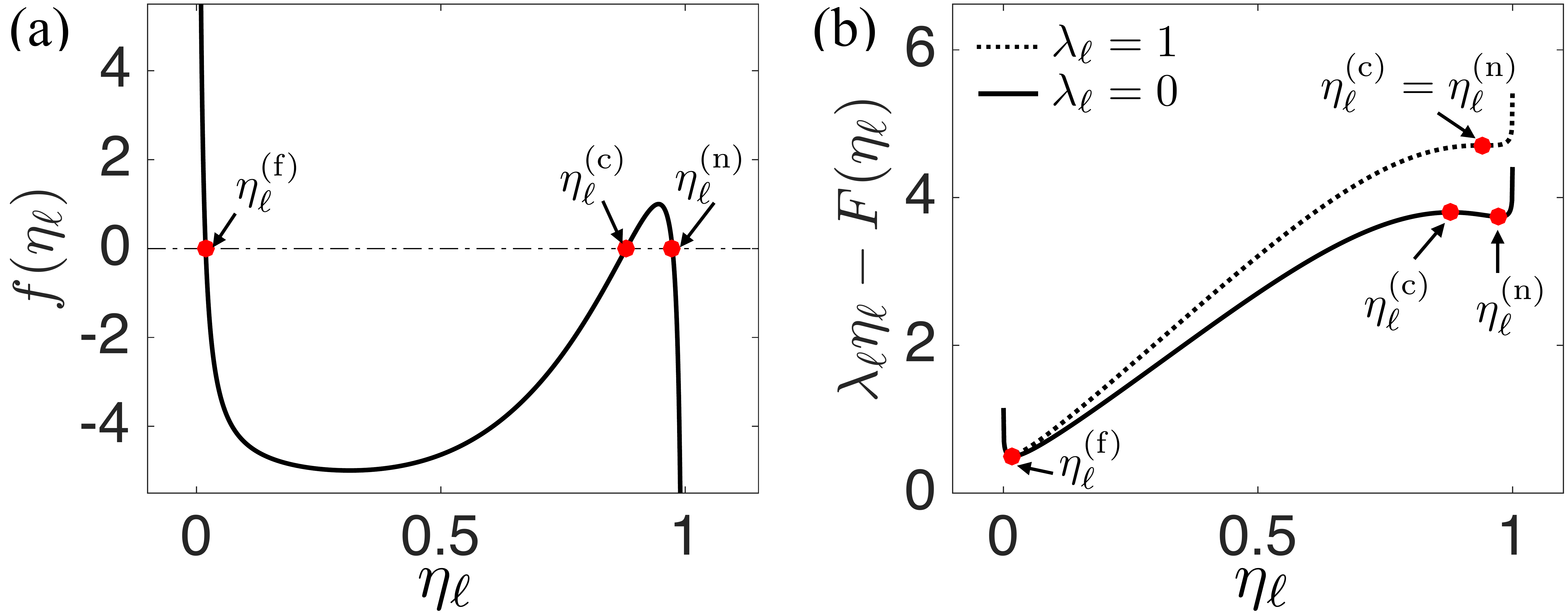}
\vspace{-0.2cm}
\caption{Line-status switch model.~(a) Function $f(\eta_\ell)$ for $a=10$, 
whose roots are the equilibrium points of Eq.~\eqref{eq:eq1} when $\lambda_\ell=0$
 (for other values of $\lambda_\ell$, see  Fig.~S1 in Supplemental Material~\cite{SM}).~(b) 
Potential function $\phi(\eta_\ell)=\lambda_\ell\eta_\ell-F(\eta_\ell)$, 
where the local minima for  $\lambda_\ell=0$ correspond to the stable equilibria in (a).~When 
$\lambda_\ell$  is increased past $1$, the local minimum $\eta_\ell^{\text{(n)}}$  merges with $\eta_\ell^{\text{(c)}}$ and then disappears.\
\label{fig1}}
\vspace{-0.2cm}
\end{figure}

Here we define $f(\eta_\ell)=a^{-1}\big[{\eta_\ell}^{-1}-(1-\eta_\ell)^{-1}\big]+a\eta_\ell^4-b$,
 where $a$ and $b$ are positive parameters.
The terms ${\eta_\ell}^{-1}$ and $-(1-\eta_\ell)^{-1}$ constrain $\eta_\ell$ above $0$ and below $1$, respectively, 
as they ensure that $f(\eta_\ell)\rightarrow\infty$ for $\eta_\ell\rightarrow0^+$ and  $f(\eta_\ell)\rightarrow-\infty$ for $\eta_\ell\rightarrow1^-$.
The term $\eta_\ell^4$ allows $f$ to have three roots---corresponding to $\eta_\ell^{\text{(f)}}$, $\eta_\ell^{\text{(c)}}$, and $\eta_\ell^{\text{(n)}}$ 
for $\lambda_\ell=0$, as shown in Fig.~\ref{fig1}(a).
The parameters $a$ and $b$ are adjustable to set $\eta_\ell^{\text{(f)}}$ close to $0$, to set $\eta_\ell^{\text{(c)}}$ and $\eta_\ell^{\text{(n)}}$ 
sufficiently  close to $1$, and to set the local maximum of $f$  to $1$.
For this choice of function $f$, Eq.~\eqref{eq:eq1} satisfies conditions (I)-(III).  Moreover,
the equation
can be rewritten as a gradient system 
$\dot{\eta}_\ell=-d\phi(\eta_\ell)/d\eta_\ell$, where
$\phi(\eta_\ell)=\lambda_\ell\eta_\ell-F(\eta_\ell)$,  and $dF(\eta_\ell)/d\eta_\ell=f(\eta_\ell)$.  
As shown in Fig.~\ref{fig1}(b), the stable equilibria of 
 this system
correspond to the local minima of  $\phi(\eta_\ell)$.

Following a perturbation, the power flowing on 
transmission  lines can change dynamically. 
When the flow on line $\ell$ reaches its capacity 
($\lambda_\ell\ge1$), 
the  system will experience a saddle-node bifurcation and the status variable $\eta_\ell$ will evolve to the stable equilibrium $\eta_\ell^{\text{(f)}}$, representing a line switch-off operation.   
This is a one-way action, since  the equilibrium $\eta_\ell^{\text{(f)}}$ is stable for any value of $\lambda_\ell$.

Having defined the dynamics of the status variables, 
we now incorporate the system's protective response
into the dynamical equations governing the state of 
the network. In a 
network of $n$ nongenerator nodes,   
each such node is an electric point where power is extracted by a load, received from  
generators, and/or redistributed among transmission lines. 
We denote by $n_\text{g}$ the number of generators, 
and by
$n_{l}$
the number of  transmission lines.
To proceed, we consider
 the extended representation of  the network~\cite{nishikawa2015comparative}
 in which each generator is now 
an additional node connected to the network through a virtual 
line  (not included in  $n_{l}$ 
and not subject to failure),
 leading to a  network of $n+n_\text{g}$ nodes. 
For notational convenience,  we reindex the generators as the first $n_\text{g}$ nodes. 

Assuming that the  voltage satisfies $|V_i|\approx1$ (in per unit)
for all nodes and that no real power is lost on transmission lines, 
we can define the state of a power system as  
$\mathbf{x}=(\pmb{\omega},\pmb{\delta},\pmb{\eta})$.  
Here, $\pmb{\omega}=(\omega_i)$ are the frequencies of the generators relative to the system's nominal frequency,
$\pmb{\delta}=(\delta_i)$ are the phase angles of all other nodes 
relative 
to a 
reference node (taken to be $i=1$, so that $\delta_1\equiv 0$), and $\pmb{\eta}=(\eta_\ell)$ are the status variables of the 
(nonvirtual)
transmission lines $\mathcal{L}$, where  $\ell\in\mathcal{L}$. The state of the system is suitably determined by the following equations:
\begin{widetext}
\begin{equation}
\label{eq:eq2}
\begin{array}{llll}
\vspace{.05cm}
&  
\dot{\omega}_i
= -\dfrac{D_i}{M_i} \omega_i - \dfrac{1}{M_i}\Big[ P_i +\sum_{j=n_\text{g}+1}^{n_\text{g}+n} \widetilde{B}_{ij} \sin \delta_{ij}\Big],   & &i= 1,2,\cdots,n_\text{g}, \\
\vspace{.1cm}
& 
\dot{\delta}_i
= \omega_i - \omega_1,   & &i= 2,\cdots,n_\text{g},\\
\vspace{.05cm}
& 
\dot{\delta}_i
= -\dfrac{1}{T_i}\Big[P_i + \sum_{j=1}^{n_\text{g}} \widetilde{B}_{ij} \sin \delta_{ij}+ \sum_{j=n_\text{g}+1}^{n_\text{g}+n} \widetilde{B}_{ij} \eta_{\ell_{i\text{-}j}} \sin \delta_{ij}\Big]-\omega_1, & &i= n_\text{g}+1,\cdots,n_\text{g}+n,\\
\vspace{.05cm}
& 
\dot{\eta}_{\ell_{i\text{-}j}}
= 10 \Big[ f(\eta_{\ell_{i\text{-}j}}) - \dfrac{\widetilde{B}_{ij}(1-\cos \delta_{ij})}{W_{\ell_{i\text{-}j}}} \Big], & & \ell_{i\text{-}j} \in \mathcal{L} .
 \end{array}
\end{equation}
\end{widetext}
Here, $\delta_{ij}=\delta_{i}-\delta_{j}$ and $\widetilde{B}$ is a 
symmetric
matrix with
nonzero elements
$\widetilde{B}_{ij}=-1/x_{\ell_{i\text{-}j}}$, where $x_{\ell_{i\text{-}j}}$ is the transient reactance of a generator or is the reactance of a transmission line,
depending on whether the line connecting $i$ and $j$  is virtual or not.  
The first two equations 
 are the swing equations describing the dynamics of the generators, 
where $M_i$ is the generator rotor inertia, $D_i$ is the rotor damping ratio, 
 and $P_i$ is the negative of the mechanical power input $P_i^{\text{(m)}}$ of the generator~\cite{swing_eq_details}.
The third equation describes
loads (and nongenerator nodes in general, under the assumption that they include some frequency-dependent power exchange) as first-order rotors, 
where $T_i$ is the load frequency ratio and $P_i$ is the power $P_i^{\text{(d)}}$ 
demanded at the node.
We further assume that $\sum_{i=1}^{n_\text{g}+n}P_i=0$, so that there exists an equilibrium point at $\omega_i=0$ and $\delta_i=cte$.
Note that the term representing the power flow on line $\ell_{i\text{-}j}$ 
is multiplied by the
status variable $\eta_{\ell_{i\text{-}j}}$, which automatically turns off the line 
in the event of an overload 
(when $\eta_\ell\rightarrow\eta_\ell^{\text{(f)}}$).
The last equation describes the dynamics of the status variables, 
where $\lambda_{\ell_{i\text{-}j}}$ in Eq.~\eqref{eq:eq1} is replaced by 
 $\widetilde{B}_{ij}(1-\cos \delta_{ij})$, the reactance energy stored in
the transmission line $\ell_{i\text{-}j}$, divided by $W_{\ell_{i\text{-}j}}$, the maximum reactance energy that line $\ell_{i\text{-}j}$ can hold. 
 The prefactor $10$ in this equation assures that the time scale for
 line failures is much shorter than that of the other dynamical changes in the network.
 For more details on the derivation of Eq.~\eqref{eq:eq2}, see Supplemental Material~\cite{SM}.

Importantly, we can show that  Eq.~\eqref{eq:eq2} can be derived from a Hamiltonian-like system of the form
\begin{equation}   
 \dot{\mathbf{x}}
 = J  \nabla\Psi(\mathbf{x}),  
\label{hamil}
\end{equation}
where $\Psi(\mathbf{x})$ is an energy function defined as
\begin{eqnarray}
\label{eq:eq4}
\Psi (\mathbf{x})
\! &= \sum_{i=1}^{n_\text{g}}\!\!\Big[ \dfrac{1}{2}M_i \omega_i^2 + 
\sum_{j=n_\text{g}+1}^{n_\text{g}+n} \widetilde{B}_{ij}(1- \cos \delta_{ij})\Big] \nonumber\\
\vspace{.06cm}
&+\sum_{i=n_\text{g}+1}^{n_\text{g}+n}\sum_{j=i+1}^{n_\text{g}+n} \widetilde{B}_{ij}(1-\cos \delta_{ij})\eta_{\ell_{i\text{-}j}}\nonumber\\
\vspace{.06cm}
&+\sum_{i=2}^{n_\text{g}+n}P_i \delta_i-\sum_{\ell_{i\text{-}j} \in \mathcal{L}} W_{\ell_{i\text{-}j}}F(\eta_{\ell_{i\text{-}j}}),
\end{eqnarray}
and  $J$ is a  
matrix of the form 
\begin{equation}
\label{eq:eq5}
J = \left[
\begin{array}{c c c c} 
J_{11} & J_{12} & J_{13} & \mathbf{0}\\
\vspace{.05cm}
-J_{12}^{T} & \mathbf{0}\ & \mathbf{0} & \mathbf{0}\\
-J_{13}^{T} & \mathbf{0} & J_{33} & \mathbf{0}\\
\mathbf{0} & \mathbf{0} & \mathbf{0} &J_{44}\\
\end{array}
\right].
\end{equation}
In this matrix, the off-diagonal blocks are 
\begin{equation}
\label{eq:eq6}
  J_{12}\! =\!\!
  \begin{bmatrix}
  \vspace{.05cm}     		
\frac{1}{M_1}& \frac{1}{M_1}&  \cdots &\! \frac{1}{M_1} \\ 
\frac{-1}{M_2} &  0  &\cdots &\! 0 \\	
0 & \frac{-1}{M_3} &  \cdots &\! 0 \\	
\vdots& \vdots & \ddots &\! \vdots\\
0 &0& \cdots &\! \frac{-1}{M_{n_\text{g}}}
  \end{bmatrix}\! , \; 
  J_{13}\! =\!\!
  \begin{bmatrix}
  \vspace{.05cm}
    \frac{1}{M_1} & \cdots &\!\! \frac{1}{M_1} \\
    0 & \cdots &\!\! 0\\
    \vdots & \vdots &\!\! \vdots\\
    0 & \cdots &\!\! 0\\
  \end{bmatrix}\! ,
\end{equation}
and the diagonal blocks are 
\begin{equation} 
\label{eq:eq7}
\begin{array}{crl}
\vspace{.05cm}
&J_{11}  &=\text{diag}(-\frac{M_1}{D_1^2},-\frac{M_2}{D_2^2},\cdots,-\frac{M_{n_\text{g}}}{D_{n_\text{g}}^2}),\\
\vspace{.05cm}
&J_{33}  &=\text{diag}(-\frac{1}{T_{n_\text{g}+1}},-\frac{1}{T_{n_\text{g}+2}},\cdots,-\frac{1}{T_{n_\text{g}+n}}),\\
& J_{44} &=10 \times \text{diag}(-\frac{1}{W_1},-\frac{1}{W_2},\cdots,-\frac{1}{W_{n_{l}}}).
\end{array}
\end{equation}
For details on the derivation of Eq.~\eqref{hamil}, see Supplemental Material~\cite{SM}.

Crucially, the matrix $J$ is the sum of a skew-symmetric matrix and a diagonal matrix with nonpositive elements, from which we can show that 
$d\Psi(\mathbf{x}(t))/dt=\nabla\Psi(\mathbf{x})^T\dot{\mathbf{x}}=\nabla\Psi(\mathbf{x})^TJ\nabla\Psi(\mathbf{x})\leq0$.
Moreover, because $J$ is also full rank
(which follows from its reduced row echelon form), 
we have that
$d\Psi(\mathbf{x}(t))/dt=\mathbf{0}$ if and only if $\nabla\Psi(\mathbf{x})=\mathbf{0}$,  
 and hence if and only if 
$\dot{\mathbf{x}}=J\nabla\Psi(\mathbf{x})=\mathbf{0}$. 
Thus, when 
the network is perturbed,
the energy function $\Psi(\mathbf{x})$ monotonically decreases as the system evolves, 
and becomes constant again only when the system reaches an equilibrium point of Eq.~\eqref{hamil} [and hence 
of Eq.~\eqref{eq:eq2}].
Such equilibria
represent stable steady  states, 
where the generators are synchronized [$\omega_1(t)=\omega_2(t)=\cdots=\omega_{n_\text{g}}(t)$],
the angle differences are fixed for all pairs of nodes, and the flow is below capacity for all operating transmission lines.

 \begin{figure}[t!] 
\includegraphics[width=.47\textwidth]{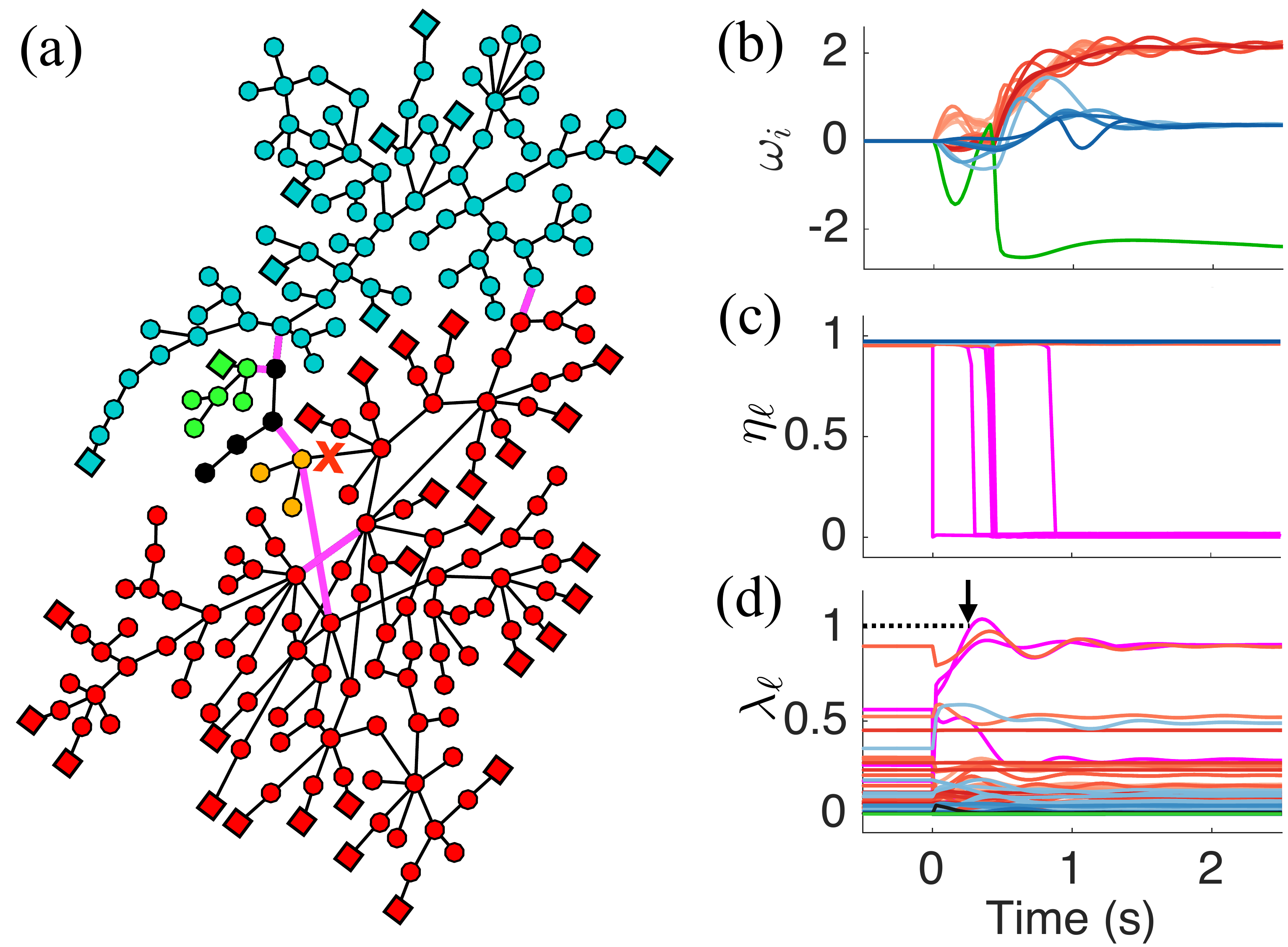}   
\caption{Simulated cascade event in 
Iceland's  
power grid.~(a) Diagram of the network, which consists of $35$ generators~($\scriptscriptstyle\square$), $189$ nongenerator nodes ($\circ$), 
and $203$ transmission lines (--) \cite{power_grid_data}.~The removal of the marked line ($\times$) 
triggers a sequence of 
$6$ subsequent
 line failures (magenta) that separate the network  into $5$ clusters (color coded).~(b-c) Corresponding 
 generator frequencies  $\omega_i$  (b) and line-status variables  $\eta_\ell$ (c) as functions of time  
[color coded as in (a)].~(d) Corresponding fraction $\lambda_\ell$ of the line capacity used,  
should the line overloaded at $0.2$ s (arrow) not be  disabled.
\label{fig3}}
\end{figure}

We  first illustrate our formalism on 
Iceland's 
power-grid network, shown in  Fig.~\ref{fig3}(a) (for parameter setting, see Supplemental Material~\cite{SM}). 
The system is designed to have a stable steady state with no additional failures when any single  transmission line is 
missing  (provided the network remains connected),
which is verified
in our simulations.
We test whether such a cascade-free steady state is actually reached following the removal of a line when 
the {\it transient dynamics} between steady states represented in our model is taken into consideration.
Starting from the stable steady state determined by Eq.~\eqref{eq:eq2},  
we simulate all $68$ single-line removal perturbations that keep the network topologically connected (performed by changing 
$\eta_\ell$ to $\eta_\ell^{\text{(f)}}$). Of these, $10$ do not converge to the best available stable steady state and instead undergo 
subsequent failures 
 (Fig.~S3 in Supplemental Material~\cite{SM}).
Insights into the underlying mechanism are provided by the example shown in Figs.~\ref{fig3}(a)-\ref{fig3}(c), where a sequence of line overloads 
separates the network into $5$ clusters. As shown in Fig.~\ref{fig3}(d), the system would eventually have approached the designed
steady state with no additional failures, 
but a line overload---whose automatic switch-off 
 triggers subsequent overloads---occurs 
 before the system can reach that state.
In this  case, no feasible trajectory exists in the
phase space connecting the initial state to the steady state predicted by 
quasi-steady-state~models.
This scenario is common in general, as shown for five other systems in the $3$rd column of Table~S2  (Supplemental Material~\cite{SM}).

When the network is subject to multiple perturbations, our framework 
 shows that the cascade outcome will generally depend on the order and timing
 of the perturbations.
A natural 
measure to quantify this difference is  the size $C'$ (i.e., number of nodes) of the largest connected cluster in the postcascade stable state. 
As an illustration, we consider the following three scenarios for two-line removal perturbations: 
(i) remove line $\ell_{i_1\text{-}j_1}$ 
and then, after the stable state is reached,  remove line $\ell_{i_2\text{-}j_2}$;  
(ii)  same as in (i) but for $\ell_{i_1\text{-}j_1}$ swapped with $\ell_{i_2\text{-}j_2}$; 
(iii) remove $\ell_{i_1\text{-}j_1}$ and $\ell_{i_2\text{-}j_2}$ concurrently.  
Considering all  $2,117$ pairs of lines $(\ell_{i_1\text{-}j_1},\ell_{i_2\text{-}j_2})$ that keep 
Iceland's network connected after their removal (but not necessarily after the resulting cascading 
failures), our simulations indicate that $30.0\%$ of these perturbations lead to cascades 
in at least one of the scenarios above.  For this subset of line pairs, we obtain that:
(a) ``order matters'' in $27.9\%$ of the cases,  in that  $C'$  differs for at least one of the scenarios;
(b)  choosing between the orders in (i) and (ii) leads 
to the largest  $C'$  in  $20.8\%$ of the  cases; 
(c) (i) and (ii)  lead to equally best $C'$ 
 in $4.3\%$ of the cases; 
(d) the concurrent removal scenario (iii) trumps  (i) and (ii) in the remaining $2.8\%$ of the 
cases (for specific examples, see 
Figs.~S4 and S5 in Supplemental Material~\cite{SM}).
Similar trends are observed for all five other systems considered, as shown in Table~S2 (Supplemental Material~\cite{SM}). 
This order dependence has potential implications for control, as 
it can be exploited in proactive line removals to prevent
subsequent failures 
 (Fig.~S6 in Supplemental Material~\cite{SM}).
This reveals a
sharp contrast between processes for which order is immaterial, such as  percolation,  and the cascades  considered here. 

\begin{figure}[t!] 
\includegraphics[width=8.5cm]{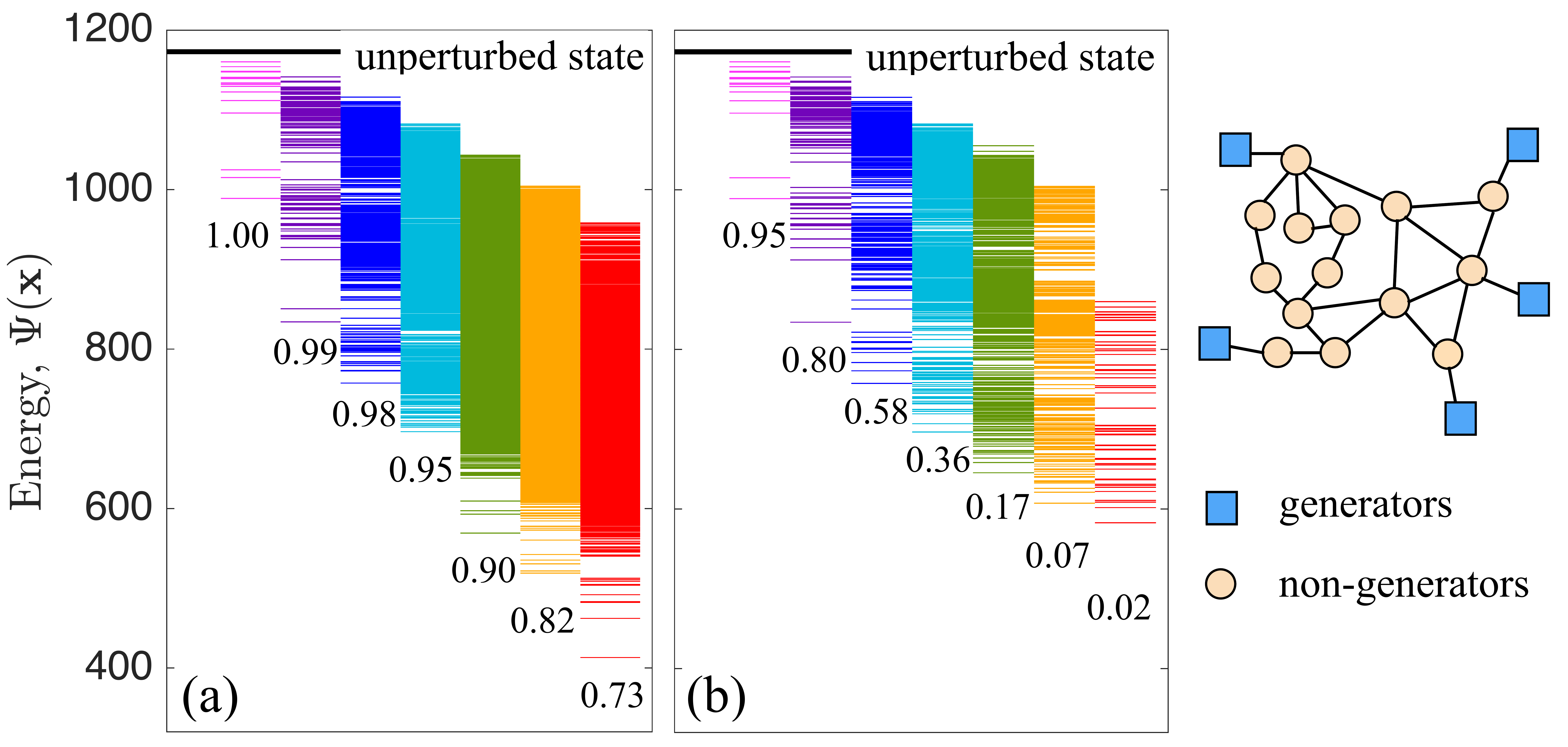}  
\caption{Energy levels  $\Psi(\mathbf{x})$ of the stable states in the $14$-bus test system.~Each panel shows all 
combinations of 1 (left column) to 7 (right) 
successive line removals that leave the network connected.~(a) All stable states  without additional failures determined using  the  
MATLAB function {\it fsolve}.~(b) Subset of stable states in (a) that the system actually evolves to for the same line 
removals as in (a).~Also marked are the fractions of perturbations for which a stable state is identified (a) and the 
fractions of those stable states actually 
reached (b).~The diagram on the rhs~shows the topology of the network.   
\label{fig5}}
\end{figure}

Taking the analysis
one step further, our formalism offers unique insight into the relation between line removal perturbations and energy levels. Figure~\ref{fig5}(a) shows all energy levels for stable steady states of the IEEE 14-bus  test system (chosen in place of 
Iceland's 
network to avoid a cluttered picture) for all combinations of $1$ to $7$ line removals that keep the network connected. Figure~\ref{fig5}(b) shows the states that the system  
actually approaches following these successive line removals---the missing states [compared to Fig.~\ref{fig5}(a)] are the ones not reached
because the system undergoes a cascade.  

Two major results follow from this.  First, it confirms that upon perturbation the system often does not reach the available stable steady state with
smallest number of failures (e.g., for 7 line removals, this is so for $98\%$ of all cases).
Second, the range of energy levels with $k+1$ line removals overlaps with the range for $k$ line removals. There are, for example, stable steady states with only one line failure at lower energy than many stable states with $2, 3, \, ..., 6$ line failures.  
This shows that, following a perturbation that could eventually lead to a stable state with multiple failures, the system can in principle be steered to a lower-energy state which has, nevertheless, a reduced number of failures. Crucially, this is possible without an external input of energy,  as the system tends to go spontaneously to lower-energy states following a perturbation.  

In summary, the model 
 presented here accounts---in a single phase space---for the interaction between the full dynamics 
 of a cascade (including transients) and the changes to the underlying network structure imposed by the resulting failures. 
The results explain
 the combinatorial impact of perturbations, identify
conditions under which a cascade may develop despite the presence of a stable state that would withstand the perturbation,
and suggest new opportunities for cascade control.

\begin{acknowledgments}
The authors thank Takashi Nishikawa for feedback on the manuscript. 
This work was supported by an ISEN Booster Award,  NSF Grant DMS-1057128, Simons Foundation Award 342906, and ARPA-E Award DE-AR0000702.
The views and opinions of authors expressed herein do not necessarily state or reflect those of the United States Government or any agency thereof.
\end{acknowledgments}

\newpage
\onecolumngrid
\clearpage

\setcounter{page}{1}
\renewcommand{\theequation}{S\arabic{equation}}
\setcounter{equation}{0}
\renewcommand{\thefigure}{S\arabic{figure}}
\setcounter{figure}{0}
\renewcommand{\thetable}{S\arabic{table}}
\setcounter{table}{0}

\begin{center}
{\bf\sc\LARGE Supplemental Material}

\bigskip
{\it Cascading Failures as Continuous Phase-Space Transitions

Yang Yang and Adilson E. Motter}
\end{center}

\section{Derivation of Equation~(\ref{eq:eq2}) in the main text}

\noindent
In a power-grid network, we define a nongenerator node as a bus  and transmission lines as electrical connections, including transformers, between pairs of such nodes. 
We 
denote the number of nongenerator nodes by
$n$, 
 the number of generators by $n_\text{g}$, and the number of transmission lines by  $n_{l}$. 
The operating condition of
the network can be characterized by the complex electrical power $S^\text{(e)}_i = P^\text{(e)}_i + \text{j}\, Q^\text{(e)}_i$ at each node $i$.
In
steady state,
all generators in an alternating current network run at the same frequency and the distribution of power flows through the network is  
determined by the complex voltage
$V_i = |V_i|e^{\text{j}\delta_i}$ at each node $i$, where $\delta_i$ is the voltage angle in the reference frame. This is determined through the power flow equations:
\begin{eqnarray}
\label{eqn:power_flow}
P^\text{(e)}_i & = &  \sum_{j=1}^{n}|V_i ||V_j |(G_{ij}\cos\delta_{ij}+B_{ij} \sin \delta_{ij}), \\
Q^\text{(e)}_i & = & -\sum_{j=1}^{n}|V_i ||V_j|(G_{ij} \sin \delta_{ij}-B_{ij} \cos \delta_{ij}),
\label{eqn:power_flow2}
\end{eqnarray}   
where $\delta_{ij}=\delta_i-\delta_j$ and $Y_{ij}=G_{ij}+\text{j}\, B_{ij}$ 
define
a Laplacian-like matrix. In this matrix, an off-diagonal element $Y_{ij}$ is the negative of the admittance of the line $\ell_{i\text{-}j}$ that connects nodes $i$ and $j$. 
Assuming that $|V_i| \approx 1$ p.u.\ for all nodes and that there is no real power lost on the transmission lines (i.e., $G_{ij}=0$), we can rewrite the real power as $P^\text{(e)}_i = \sum_{j=1}^{n} B_{ij}  \sin (\delta_i -\delta_j)$, where $B$ is a symmetric matrix with off-diagonal elements $B_{ij}=-1/x_\ell$ and  $x_\ell$ is the reactance of line $\ell$. These assumptions are valid throughout this paper.

In general, the state of both generators and loads can change in time. 
During a disturbance, the generator rotors decelerate or accelerate with respect to the nominal frequency ($60$ Hz in the U.S. and $50$ Hz in Europe, including Iceland). The dynamics of the generator rotor is governed by the swing equation:
$M_i \dfrac{d\omega_i}{dt} + D_i \omega_i = P^\text{(m)}_i - P^\text{(e)}_i$,
where $\omega_i$ is the  frequency (relative to the nominal frequency), $M_i$ is  the rotor inertia, $D_i$ is the rotor damping ratio, and $P^\text{(m)}_i$ is  
the net shaft power input into the generator.  
Considering Eq.~(\ref{eqn:power_flow}), we can combine the power flow equations into the swing equation as 
\begin{equation}
\label{eqn:swing_eqn}
M_i \dfrac{d\omega_i}{dt} + D_i \omega_i = P^\text{(m)}_i -\sum_{j=1}^{n} B_{ij} \sin (\delta_i -\delta_j).
\end{equation}
Here we choose the angle of the first generator as the  reference angle, and we define $\delta_i = \alpha_i-\alpha_1$, where $\dfrac{d\alpha_i}{dt}=\omega_i$. In principle, the power consumed by load nodes could depend nonlinearly on the frequency at that node.  Assuming that the frequency at each node does not deviate strongly from the nominal frequency,   
we can use a linearized power-frequency relation to describe the dynamics of the node connected with a load:
\begin{equation}  
\label{eqn:load_freq}
T_i \dfrac{d\delta_i}{dt}=P^\text{(d)}_i -\sum_{j=1}^{n} B_{ij} \sin (\delta_i -\delta_j)-\omega_1,
\end{equation}
where $T_i$ is a positive constant and $P^\text{(d)}_i$ is the power requested by the load.

To account for the internal reactances of the generators,  
we adopted an extended representation of the power grid. As explained in the main text,  in this extended representation we add $n_\text{g}$ nodes  connected to the network through virtual lossless lines to represent the  $n_\text{g}$ generators in the system (see Fig.~\ref{fig2}).  The reactance of a  virtual line represents the transient reactance of the corresponding generator.  Accordingly, we modify the $B$ matrix in Eqs.~(\ref{eqn:swing_eqn}) and (\ref{eqn:load_freq}) as 
\begin{equation}
\widetilde{B} = 
	\begin{bmatrix}
		0 & X \\
		X^{T} & B
	\end{bmatrix},
\label{eqn:Bij}
\end{equation}
where $X_{ij}$ is the reciprocal of the transient reactance of the $i$th generator that connects the $j$th node in the network. Combining together the dynamics of generators and loads, the equations of motion take the following form:
\begin{equation}
\label{eqn:sp}
\begin{array}{rlll}
M_i \dfrac{d\omega_i}{dt} + D_i \omega_i &= P^\text{(m)}_i -\sum_{j=1}^{n_\text{g}+n} \widetilde{B}_{ij} \sin (\delta_i -\delta_j),   & &i= 1,2,\cdots,n_\text{g}, \\ [2mm]
\dfrac{d\delta_i}{dt} &= \omega_i - \omega_1,   & &i= 2,\cdots,n_\text{g},\\[3mm]
T_i \dfrac{d\delta_i}{dt}&=-P^\text{(d)}_i - \sum_{j=1}^{n_\text{g}+n} \widetilde{B}_{ij} \sin (\delta_i -\delta_j)-\omega_1, & &i= n_\text{g}+1,\cdots,n_\text{g}+n,
 \end{array}
\end{equation}
where, for simplicity,  the last equation is assumed to apply to all nongenerator nodes under the assumption that they all include some frequency-dependent power exchange.
Adding up all the equations in Eq.~(\ref{eqn:sp}), at the fixed point of the dynamics 
(where $d\omega_i/dt=0$, $d\delta_i/dt=0$) we obtain $(\sum_{i=1}^{n_\text{g}}D_i+n)\omega_1=\sum_{i=1}^{n_\text{g}}{P^{\text{(m)}}_i}-\sum_{n_\text{g}+1}^{n_\text{g}+n}P^{\text{(d)}}_i$. We assume that
the real power is balanced, i.e., $\sum_{i=1}^{n_\text{g}}{P^{\text{(m)}}_i}=\sum_{n_\text{g}+1}^{n_\text{g}+n}P^{\text{(d)}}_i$, and hence 
all generators operate with the nominal frequency at the fixed point.

To complete the derivation of Eq.~(\ref{eq:eq2}), for each transmission line $\ell_{i\text{-}j}$ we need to define the fraction $\lambda_{\ell_{i\text{-}j}}$ 
of the line capacity used by the flow.
This quantity is determined by the average amount of reactive power stored in the transmission line, calculated as $\dfrac{1}{2}|(V_i-V_j)I_{ij}^*|$,
where $I_{ij}= \text{j}\, B_{ij}(V_i-V_j)$ is the current on the line. Noting that  $|V_i|\approx |V_j|$  (both approximately equal to $1$ p.u.), 
we can determine this reactive energy  as  $\widetilde{B}_{ij}(1-\cos \delta_{ij})$,  and
replace $\lambda_{\ell_{i\text{-}j}}$ in Eq.\ (\ref{eq:eq1}) by this reactive energy divided by $W_{\ell_{i\text{-}j}}$ (the maximum reactive energy that line $\ell_{i\text{-}j}$ can hold). 
By incorporating the dynamics of the status variables, as presented in the main text,  into the dynamics of the power system in Eq.~(\ref{eqn:sp}), we finally obtain Eq.~(\ref{eq:eq2}). This equation describes, at the same time, the state variables of the power system and the status of the transmission lines.

\section{Proof of Equation~(\ref{hamil}) in the main text}

We now explicitly show the equivalence between Eqs.~(\ref{hamil})  and (\ref{eq:eq2}) in the main text. We recall that the state of a power system is defined as 
$\mathbf{x}=(\pmb{\omega}, \pmb{\delta},\pmb{\eta})$.  
As in the main text, the (relative) frequencies of the generators are represented by a vector $\pmb{\omega}$ of size $ n_\text{g}\times 1$,
the phase angles of all nodes relative to the reference node are represented by a vector $\pmb{\delta}$ of size $(n_\text{g}+n-1)\times 1$,
and the status variables of the transmission lines are represented by a vector $\pmb{\eta}$ of size $ n_{l}\times 1$.
To facilitate our verification of Eq.~(\ref{hamil}), we further separate $\pmb{\delta}$ into $\pmb{\delta}=(\pmb{\delta^{'}},\pmb{\delta^{''}})$,
where the vector $\pmb{\delta}^{'}=(\delta_2,\cdots, \delta_{n_\text{g}})$ represents the phase angles of the generator nodes and 
vector $\pmb{\delta}^{''}=(\delta_{n_\text{g}+1}, \cdots, \delta_{n_\text{g}+n})$ represents the phase angles of nongenerator nodes.

According to  Eq.~(\ref{eq:eq4}), the gradient of $\Psi(\mathbf{x})$ can be decomposed as
$\nabla\Psi(\mathbf{x})=\big[ \nabla_{\pmb{\omega}} \Psi(\mathbf{x}), \nabla_{\pmb{\delta}^{'}}\Psi(\mathbf{x}), \nabla_{\pmb{\delta^{''}}}\Psi(\mathbf{x}), \nabla_{\pmb{\eta}}\Psi(\mathbf{x}) \big]$, where
\begin{align*}  
\nabla_{\pmb{\omega}} \Psi(\mathbf{x}) &= 
\begin{bmatrix}
 M_1 \omega_1, M_2 \omega_2,\cdots,\,M_{n_\text{g}} \omega_{n_\text{g}}
 \end{bmatrix}^{T}, \stepcounter{equation}\tag{\theequation}\\[5mm]
\nabla_{\pmb{\delta^{'}}} \Psi(\mathbf{x}) &= 
\begin{bmatrix}
\sum_{j=1}^{n_\text{g}+n} \widetilde{B}_{2j} \sin \delta_{2j}+P_2\\[2mm]
\sum_{j=1}^{n_\text{g}+n} \widetilde{B}_{3j} \sin \delta_{3j}+P_3\\[2mm]
\vdots\\[2mm]
\sum_{j=1}^{n_\text{g}+n} \widetilde{B}_{n_\text{g}j} \sin \delta_{n_\text{g}j}+P_{n_\text{g}}
\end{bmatrix}, \stepcounter{equation}\tag{\theequation}\\[5mm]
\nabla_{\pmb{\delta^{''}}} \Psi(\mathbf{x}) &= 
\begin{bmatrix}
\sum_{j=1}^{n_\text{g}} \widetilde{B}_{(n_\text{g}+1)j} \sin \delta_{(n_\text{g}+1)j}+\sum_{j=n_\text{g}+1}^{n_\text{g}+n} \widetilde{B}_{(n_\text{g}+1)j} \sin \delta_{(n_\text{g}+1)j}\eta_{\ell_{(n_\text{g}+1)\text{-}j}}+P_{n_\text{g}+1}\\[2mm]
\sum_{j=1}^{n_\text{g}} \widetilde{B}_{(n_\text{g}+2)j} \sin \delta_{(n_\text{g}+2)j}+\sum_{j=n_\text{g}+1}^{n_\text{g}+n} \widetilde{B}_{(n_\text{g}+2)j} \sin \delta_{(n_\text{g}+2)j}\eta_{\ell_{(n_\text{g}+2)\text{-}j}}+P_{n_\text{g}+2}\\[2mm]
\vdots\\[2mm]
\sum_{j=1}^{n_\text{g}} \widetilde{B}_{(n_\text{g}+n)j} \sin \delta_{(n_\text{g}+n)j}+\sum_{j=n_\text{g}+1}^{n_\text{g}+n} \widetilde{B}_{(n_\text{g}+n)j} \sin \delta_{(n_\text{g}+n)j}\eta_{\ell_{(n_\text{g}+n)\text{-}j}}+P_{n_\text{g}+n}\\[2mm]
\end{bmatrix},\stepcounter{equation}\tag{\theequation} \\
\nabla_{\pmb{\eta}}\Psi(\mathbf{x}) &=
\begin{bmatrix}
\vdots\\
 \widetilde{B}_{ij}(1- \cos \delta_{ij}) - W_{\ell_{i\text{-}j}}f(\eta_{\ell_{i\text{-}j}})\stepcounter{equation}\tag{\theequation}\\[2mm]
 \vdots\\[2mm]
\end{bmatrix}.
\end{align*}
In the 
last equation,
each component of $\nabla_{\pmb{\eta}}\Psi(\mathbf{x})$ corresponds to the derivative of $\Psi(\mathbf{x})$  with respect to the status variable of a transmission line $\ell_{i\text{-}j}$. 
For notational convenience, in the expression
of
$\nabla_{\pmb{\delta^{''}}}\Psi(\mathbf{x})$
we introduce the constants $\eta_{\ell_{(n_\text{g}+i)\text{-}j}}\equiv 0$ associated with pairs of nodes $\big( n_\text{g}+i,j\big)$ that are not connected by a transmission line (these constants should not be confused with the components of $\pmb{\eta}$, which are variables associated with  pairs of nodes that are connected).

Making use of the definition of matrix $J$ in Eqs.~(\ref{eq:eq5})-(\ref{eq:eq7}) of  the main text, we obtain
\begin{equation}
\label{eq:sm8}
\begin{array}{cr}
&J_{11}\nabla_{\pmb{\omega}}\Psi(\mathbf{x})+J_{12}\nabla_{\pmb{\delta^{'}}}\Psi(\mathbf{x})+J_{13}\nabla_{\pmb{\delta^{''}}}\Psi(\mathbf{x})
=
\begin{bmatrix}
\frac{-D_1}{M_1}\omega_1\\[2mm]
\frac{-D_2}{M_2}\omega_2\\[2mm]
\vdots\\[2mm]
\frac{-D_{n_\text{g}}}{M_{n_\text{g}}}\omega_{n_\text{g}}
\end{bmatrix}
+
\begin{bmatrix}
\frac{1}{M_1}\sum_{i=2}^{n_\text{g}}(\sum_{j=1}^{n_\text{g}+n} \widetilde{B}_{ij} \sin \delta_{ij}+P_i)\\[2mm]
\frac{-1}{M_2}(\sum_{j=1}^{n_\text{g}+n} \widetilde{B}_{2j} \sin \delta_{2j}+P_2)\\[2mm]
\vdots\\[2mm]
\frac{-1}{M_{n_\text{g}}}(\sum_{j=1}^{n_\text{g}+n} \widetilde{B}_{2j} \sin \delta_{{n_\text{g}j}}+P_{n_\text{g}})
\end{bmatrix}~~ ~~~~~~~~\\[15mm]
&+
\begin{bmatrix}
\frac{1}{M_1}\sum_{i=n_\text{g}+1}^{n_\text{g}+n}(\sum_{j=1}^{n_\text{g}} \widetilde{B}_{ij} \sin \delta_{ij}+\sum_{j=n_\text{g}+1}^{n_\text{g}+n} \widetilde{B}_{ij} \sin \delta_{ij}\eta_{\ell_{i\text{-}j}}+P_i)\\[5mm]
0\\[2mm]
\vdots\\[2mm]
0
\end{bmatrix}.~~~~~~~~
\end{array}
\end{equation}
We note that $\sum_{i=n_\text{g}+1}^{n_\text{g}+n}\sum_{j=n_\text{g}+1}^{n_\text{g}+n}\widetilde{B}_{ij} \sin \delta_{ij}\eta_{\ell_{i\text{-}j}}=0$ 
given that $\widetilde{B}_{ij}=\widetilde{B}_{ji}$ and $\sin \delta_{ij}=-\sin \delta_{ji}$. Accordingly, the first row of the rhs of Eq.~(\ref{eq:sm8}) can be rewritten as
\begin{equation}
\begin{array}{ll}
&-\dfrac{D_1}{M_1}\omega_1+\dfrac{1}{M_1}(\sum_{i=2}^{n_\text{g}}\sum_{j=1}^{n_\text{g}+n} \widetilde{B}_{ij} \sin \delta_{ij}
+\sum_{i=n_\text{g}+1}^{n_\text{g}+n}\sum_{j=1}^{n_\text{g}} \widetilde{B}_{ij} \sin \delta_{ij})+\dfrac{1}{M_1}\sum_{i=2}^{n_\text{g}+n}P_i\\[3mm]
=& -\dfrac{D_1}{M_1}\omega_1+
\dfrac{1}{M_1}(\sum_{i=2}^{n_\text{g}}\sum_{j=1}^{n_\text{g}+n} \widetilde{B}_{ij} \sin \delta_{ij}
+\sum_{i=n_\text{g}+1}^{n_\text{g}+n}\sum_{j=1}^{n_\text{g}} \widetilde{B}_{ij} \sin \delta_{ij})-\dfrac{1}{M_1}P_1\\[3mm]
=& -\dfrac{D_1}{M_1}\omega_1+
\dfrac{1}{M_1}(\sum_{i=2}^{n_\text{g}}\sum_{j=1}^{n_\text{g}+n} \widetilde{B}_{ij} \sin \delta_{ij}
+\sum_{j=n_\text{g}+1}^{n_\text{g}+n}\sum_{i=1}^{n_\text{g}} \widetilde{B}_{ji} \sin \delta_{ji})-\dfrac{1}{M_1}P_1\\[3mm]
=& -\dfrac{D_1}{M_1}\omega_1+
\dfrac{1}{M_1}(\sum_{i=2}^{n_\text{g}}\sum_{j=1}^{n_\text{g}+n} \widetilde{B}_{ij} \sin \delta_{ij}
-\sum_{j=n_\text{g}+1}^{n_\text{g}+n}\sum_{i=1}^{n_\text{g}} \widetilde{B}_{ij} \sin \delta_{ij})-\dfrac{1}{M_1}P_1\\[3mm]
=& -\dfrac{D_1}{M_1}\omega_1+
\dfrac{1}{M_1}(\sum_{i=2}^{n_\text{g}}\sum_{j=1}^{n_\text{g}+n} \widetilde{B}_{ij} \sin \delta_{ij}
-\sum_{j=1}^{n_\text{g}+n}\sum_{i=1}^{n_\text{g}} \widetilde{B}_{ij} \sin \delta_{ij})-\dfrac{1}{M_1}P_1\\[3mm]
=& -\dfrac{D_1}{M_1}\omega_1-
\dfrac{1}{M_1}\sum_{j=1}^{n_\text{g}+n} \widetilde{B}_{ij} \sin \delta_{ij}-\dfrac{1}{M_1}P_1 \, .\\[3mm]
\end{array}
\end{equation}
In the above derivation, we assume that the total amount of power injection into the network equals the total amount of power extracted from the network (i.e., $\sum_{i=1}^{n_\text{g}+n}P_i=0$), and we use the fact that $\widetilde{B}_{ij} =0$ for $i,j=1,2,\cdots,n_\text{g}$.
Similar calculation applies to the other rows of Eq.~(\ref{eq:sm8}).
Then, noting that the rhs of Eq.~(\ref{eq:sm8}) is the same as the rhs of the first equation in~(\ref{eq:eq2}),
 we conclude that
$J_{11}\nabla_{\pmb{\omega}}\Psi(\mathbf{x})+J_{12}\nabla_{\pmb{\delta^{'}}}\Psi(\mathbf{x})+J_{13}\nabla_{\pmb{\delta^{''}}}\Psi(\mathbf{x})=\dfrac{d\pmb{\omega}}{dt}$.

Next, we  establish the equivalence between the other components of Eq.~(\ref{hamil}) and the 
other equations
 in~(\ref{eq:eq2}). We obtain the second equation  in~(\ref{eq:eq2}) using that
\begin{equation}
-J_{12}^{T}\nabla_{\pmb{\omega}}\Psi(\mathbf{x})
=
\begin{bmatrix}
\omega_2-\omega_1\\
\omega_3-\omega_1\\
\vdots\\
\omega_{n_\text{g}}-\omega_1\\
\end{bmatrix}
=\frac{d\pmb{\delta^{'}}}{dt},
\end{equation}
the third equation in~(\ref{eq:eq2}) using that
\begin{equation}
\begin{array}{ll}
-J_{13}^{T}\nabla_{\pmb{\omega}}\Psi(\mathbf{x})+J_{33}\nabla_{\pmb{\delta^{''}}}\Psi(\mathbf{x})=\\[3mm]
\begin{bmatrix}
-\omega_1\\[3mm]
-\omega_1\\[3mm]
\vdots\\[3mm]
-\omega_1\\
\end{bmatrix}
+
\begin{bmatrix}
-\frac{1}{T_{n_\text{g}+1}}\big(\sum_{j=1}^{n_\text{g}} \widetilde{B}_{(n_\text{g}+1)j} \sin \delta_{(n_\text{g}+1)j}+\sum_{j=n_\text{g}+1}^{n_\text{g}+n} \widetilde{B}_{(n_\text{g}+1)j} \sin \delta_{(n_\text{g}+1)j}\eta_{\ell_{(n_\text{g}+1)\text{-}j}}+P_{n_\text{g}+1}\big)\\[2mm]
-\frac{1}{T_{n_\text{g}+2}}(\sum_{j=1}^{n_\text{g}} \widetilde{B}_{(n_\text{g}+2)j} \sin \delta_{(n_\text{g}+2)j}+\sum_{j=n_\text{g}+1}^{n_\text{g}+n} \widetilde{B}_{(n_\text{g}+2)j} \sin \delta_{(n_\text{g}+2)j}\eta_{\ell_{(n_\text{g}+2)\text{-}j}}+P_{n_\text{g}+2})\\[2mm]
\vdots\\[2mm]
-\frac{1}{T_{n_\text{g}+n}}(\sum_{j=1}^{n_\text{g}} \widetilde{B}_{(n_\text{g}+n)j} \sin \delta_{(n_\text{g}+n)j}+\sum_{j=n_\text{g}+1}^{n_\text{g}+n} \widetilde{B}_{(n_\text{g}+n)j} \sin \delta_{(n_\text{g}+n)j}\eta_{\ell_{(n_\text{g}+n)\text{-}j}}+P_{n_\text{g}+n})\\[2mm]
\end{bmatrix}\\[3mm]
\hspace{4.5cm}=\dfrac{d\pmb{\delta^{''}}}{dt},
\end{array}
\end{equation}
and the fourth equation in  (\ref{eq:eq2}) using that
\begin{equation}
\label{eq:sm12}
J_{44}\nabla_{\pmb{\eta}}\Psi(\mathbf{x})
=
\begin{bmatrix}
\vdots\\
{\displaystyle -\frac{\widetilde{B}_{ij}(1- \cos \delta_{ij})}{W_{\ell_{i\text{-}j}}} + f(\eta_{\ell_{i\text{-}j}}) }\\[2mm]
 \vdots\\[2mm]
\end{bmatrix}=\frac{d\pmb{\eta}}{dt} \, .
\end{equation}
Combining Eqs.~(\ref{eq:sm8})-(\ref{eq:sm12}), we have proved Eq.~(\ref{hamil}) in the main text.
We note that a similar result on a $3$-bus network is reported in Ref.~\cite{supp_zheng2010bi}. However, that study 
does not offer a framework
to address a network with an arbitrary number of buses, which is derived here. 

For completeness, we note that our energy-function formulation is also naturally suited for stability analysis.  
Rewriting Eq.~\eqref{hamil} as $\dot{\mathbf{x}}= \mathbf{g}(\mathbf{x})$, if  $\mathbf{x}^{*}$ is an equilibrium state, by definition we have 
$ \mathbf{g}(\mathbf{x}^{*}) =\mathbf{0}$.
The stability of 
this state is then
 determined by the eigenvalues of the Jacobian matrix 
$d\mathbf{g}/d\mathbf{x} \rvert_{\mathbf{x}=\mathbf{x}^*}= J H(\mathbf{x}^*)$,  
where  the components of matrix $H$ are given by $H_{kk'}=\dfrac{\partial^2 \Psi}{\partial x_{k}\partial x_{k'}}$.

\section{Generalized Hamiltonian-like energy when the network splits}

Our analysis of the energy function $\Psi(\mathbf{x})$ assumes a balance of power in the network,  which is guaranteed when the network remains connected but can be violated when it splits during a cascade. This is easily remediated, however, by extending the formalism to introduce a reference generator in each cluster to mimic the system's operation  of rebalancing real power. As shown below,  the reference generator  in each cluster then serves as a slack bus  that prevents imbalances between power input and output, which  would cause acceleration or deceleration of generators.
In our applications in the paper, 
the network remains connected or else we consider each cluster separately.

We first note that the choice of the reference generator does not impact our description of power-grid dynamics, 
as long as the power input and output at each node is fixed.
Then we can show that in a power  network satisfying $\sum_{i=1}^{n_\text{g}+n}P_i=0$, the value of the energylike function $\Psi(\mathbf{x})$ remains the same for different choices of the reference node.
Specifically, when the angle of generator $1$ is chosen to be the reference, the state is originally defined as $\mathbf{x}=(\pmb{\omega},\pmb{\delta},\pmb{\eta})$, 
where the frequency is defined by  $\omega_i=\dfrac{d\alpha_i}{dt}$, 
the reference angle is defined by $\pmb{\delta}=(\delta_2,\delta_3,\cdots,\delta_{n_\text{g}}) \equiv (\alpha_2-\alpha_1,\alpha_3-\alpha_1,\cdots,\alpha_{n_\text{g}}-\alpha_1)$, 
and 
$\alpha_i$ is the phase angle of node $i$. 
Here, if we change the reference node to be node $r$, the 
state vector needs to be redefined as 
$\mathbf{x}^{(r)}=(\pmb{\omega},\pmb{\delta}^{(r)},\pmb{\eta})$, where
\begin{equation}
\pmb{\delta}^{(r)}=(\delta^{(r)}_1,\delta^{(r)}_2,\cdots,\delta^{(r)}_{r-1},\delta^{(r)}_{r+1},\cdots,\delta^{(r)}_{n_\text{g} +n}) \equiv
 (\alpha_1-\alpha_r,\alpha_2-\alpha_r,\cdots,\alpha_{r-1}-\alpha_r,\alpha_{r+1}-\alpha_r,\cdots,\alpha_{n_\text{g}+n}-\alpha_r).
\end{equation}
We note that, in the redefined phase space, the angle difference between two nodes remains the same as in the original frame, 
i.e., $\delta^{(r)}_{ij}\equiv\delta^{(r)}_i-\delta^{(r)}_j=(\alpha_i-\alpha_r)-(\alpha_j-\alpha_r)=\alpha_i-\alpha_j=\delta_i-\delta_j\equiv \delta_{ij}$.
Therefore, the energylike function in the new reference frame, which we denote  $\Psi (\mathbf{x}^{(r)})$, can be written 
using the expression in 
Eq.~(\ref{eq:eq4}) but with the term $\sum_{i=2}^{n_\text{g}+n}P_i \delta_i$   replaced by $\sum_{\substack{i=1 \\ i\neq r}}^{n_\text{g}+n}P_i \delta^{(r)}_i$.
By making use of  $\sum_{i=1}^{n_\text{g}+n}P_i=0$, we can further derive
\begin{equation}
\sum_{\substack{i=1 \\ i\neq r}}^{n_\text{g}+n}P_i \delta^{(r)}_i=
\sum_{\substack{i=1 \\ i\neq r}}^{n_\text{g}+n}P_i \alpha_i- \Big[\sum_{\substack{i=1 \\ i\neq r}}^{n_\text{g}+n}P_i \Big] \alpha_r=
\sum_{\substack{i=1 \\ i\neq r}}^{n_\text{g}+n}P_i\alpha_i+P_r\alpha_r=
\sum_{i=2}^{n_\text{g}+n}P_i\alpha_i+P_1\alpha_1=
\sum_{i=2}^{n_\text{g}+n}P_i \alpha_i- \Big[ \sum_{i=2}^{n_\text{g}+n}P_i \Big]\alpha_1=
\sum_{i=2}^{n_\text{g}+n}P_i \delta_i.
\end{equation}
We can conclude, therefore, that  the energy function remains unchanged when we choose a different  reference generator node.
In the same way we proved Eq.~(\ref{hamil}), we can also prove that 
\begin{equation}
 \dfrac{d\mathbf{x}^{(r)}}{dt}= J^{(r)}  \nabla\Psi(\mathbf{x}^{(r)}), 
\end{equation}
where $J^{(r)}$ is a full-rank matrix of the form 
\begin{equation}
J^{(r)} = \left[
\begin{array}{c c c c} 
J_{11} & J^{(r)}_{12} & J^{(r)}_{13} & \mathbf{0}\\
\vspace{.05cm}
-[J^{(r)}_{12}]^{T} & \mathbf{0}\ & \mathbf{0} & \mathbf{0}\\
-[J^{(r)}_{13}]^{T} & \mathbf{0} & J_{33} & \mathbf{0}\\
\mathbf{0} & \mathbf{0} & \mathbf{0} &J_{44}\\
\end{array}
\right].
\end{equation}
In this matrix, the off-diagonal blocks are 
\begin{equation}
  J^{(r)}_{12}\! =\!\!
  \begin{bmatrix}
  \vspace{.05cm}     		
\frac{-1}{M_1} &  0 &0  &\cdots & \! \cdots &\cdots &\cdots & \! 0 \\	
0 & \frac{-1}{M_2} &  0 & \cdots & \! \cdots &\cdots  &\cdots &\! 0 \\	
\vdots& \vdots & \ddots &\! \ddots &\!\ddots  &\ddots &\ddots &\vdots\\
\vspace{.05cm}
0 & 0 &  \cdots &\! \frac{-1}{M_{r-1}} & 0 &\! 0 & \cdots &0\\	
\vspace{.05cm}
\frac{1}{M_{r}} & \frac{1}{M_{r}} & \cdots & \frac{1}{M_{r}}  & \frac{1}{M_{r}} &  \frac{1}{M_{r}}  &\cdots & \!\frac{1}{M_{r}}\\	
0 & 0 &  \cdots &\! 0 & 0 &  \frac{-1}{M_{r+1}} &\cdots & 0 \\	
\vdots& \vdots & \ddots &\! \ddots &\!\ddots  &\ddots &\ddots &\vdots\\
0 &0& \cdots & \! \cdots &\cdots &  \cdots& \cdots & \frac{-1}{M_{n_\text{g}}}
  \end{bmatrix}\! , \;  
  J^{(r)}_{13}\! =\!\!
  \begin{bmatrix}
  \vspace{.05cm}
    0 & \cdots &\!\! 0\\
    \vdots & \vdots &\!\! \vdots\\
    0 & \cdots &\!\! 0\\
        \frac{1}{M_r} & \cdots &\!\! \frac{1}{M_r} \\
    0 & \cdots &\!\! 0\\
    \vdots & \vdots &\!\! \vdots\\
    0 & \cdots &\!\! 0\\
  \end{bmatrix}\! ,
\end{equation}
where  $J^{(r)}_{13}$ is an $n_\text{g} \times n$ matrix with nonzero elements 
in row $r$, and the diagonal blocks are the same as defined in Eq.~(\ref{eq:eq7}) of the main text.

To proceed, we consider the situation in which a network $G$ is split into $k\ge2$ disconnected clusters $G_1,
 \; \cdots, G_k$ due to line failures. 
Without loss of generality, we assume that $G_1$ is the largest cluster in the network and 
that $G_1$ contains the reference node $r$ (which can be otherwise reassigned since the choice of the reference node does not impact the dynamics 
or the value of $\Psi$). After rebalancing the power input and output in $G_1$ by setting $\sum_{i \in G_1}P_i=0$,  
we can consider the dynamics in the corresponding subspace of the phase space of the system using our formalism:
\begin{equation}
 \dfrac{d\overline{\mathbf{x}^{(r)}}}{dt}= \overline{J^{(r)}} \nabla\Psi(\overline{\mathbf{x}^{(r)}}).
\end{equation}
Here, the substate $\overline{\mathbf{x}^{(r)}}$ contains only the variables for the nodes 
(generators and nongenerators) and transmission lines in the cluster $G_1$.
The matrix $\overline{J^{(r)}}$ and  function $\Psi(\overline{\mathbf{x}^{(r)}})$ are defined on the cluster $G_1$ in the same way as 
 ${J^{(r)}}$ and $\Psi({\mathbf{x}^{(r)}})$
were defined for the entire network $G$.
The same procedure can be used in each cluster of the network, thereby leading to a self-consistent approach that can be applied to the general case in which the network splits into disconnected clusters during a cascade.

There is an exceptional case under which our formalism will fail. Recalling that $P_r$ is the negative of the mechanical power input from the generator node $r$, the sum $\sum_{\substack{i \in G_1\\i \neq r}}P_i = -P_r$  must 
be a nonnegative number  smaller than the reciprocal of 
the transient reactance
of the generator.  
The corresponding condition must hold true for each cluster.
If for any cluster  we are not able to select a reference generator 
such that this condition is satisfied,  we 
declare
it 
an unsolvable state. 
 This state corresponds to the situation in which we are not able to rebalance the power input and output in a cluster by adjusting the input from any single generator in the cluster,  
and hence the cascade will necessarily continue to propagate in that cluster. Other operations, such as  shedding  power and adjusting the input from 
multiple generators, would be generally needed in this case to 
rebalance the system.

\newpage

\section{Power network data}\
To the best of our knowledge, this 
is the first study on power grids to account for both line overload  and generator dynamics in the same cascade event. Our analysis requires static power flow 
data and dynamic data on the parameters of the generators. 
The power flow data include the graph topology and electrical parameters of the transmission lines as well as the power-demand and generator-output data.
The power-grid network of  Iceland considered in this work is the largest publicly available system known to us with consistent static and dynamic data, in which the static generator data can be matched to the generator data used in stability tests. 
This network has the additional advantage of being isolated, and 
therefore involves no assumptions on possible external connections. 
The other networks used to complement our analysis are standard test systems often employed in power flow studies. They consist of four IEEE test transmission
systems and the PEGASE 89-bus system, which represents a 
transmission network in Europe available through Matpower~5.0.

Iceland's  network is shown in Fig.~\ref{iceland_network}.
The static and dynamic parameters of this system are provided in Ref.~\cite{power_grid_data}.
The available data include the real power $P^\text{(m)}_i$  supplied by each generator, the real power $P^\text{(d)}_i$ demanded at 
each node, the reactance $x_\ell$ of each transmission line, and the dynamic data (including rotor inertia  $M_i$
and transient reactance) of each generator.
We choose the load frequency ratio to be $T_i=1$ for every nongenerator node and the rotor damping ratio to be $D_i=5$ for every generator  in the network. 
The assumption that these parameters are the same for the different nodes is not essential; other choices are possible, and are consistent with the mathematical
assumptions underlying Eq.~\eqref{eqn:sp} and our formalism,  provided
that the parameters in the denominators of Eqs.~\eqref{eq:eq6} and \eqref{eq:eq7} are nonzero.
If there are two lines connecting a pair of nodes, we replace them by a single line with the combined impedance. 
For one of the lines, whose actual reactance is not available in the data, the reactance is assigned to be $0.0001$ p.u.\ 
 to assure that this line has a large capacity. 
To define the capacity of each
transmission line, we calculate the steady-state power-flow solutions [determined by Eq.~(\ref{eqn:power_flow})] for all
possible one-line failure scenarios that keep the
network connected. We then assign the capacity $W_\ell$ of each line to be $110\%$ of the 
maximum reactance energy that the line $\ell$ stores in these solutions. In this way,  the system has at least one 
stable steady state
 when a transmission line is disconnected (as long as the network remains connected, as assumed).

The basic properties of the test systems are listed in Table \ref{table:results_comparison}~\cite{milano2005open}, where we also include 
Iceland's network for completeness.
In particular, the IEEE $14$-bus test system  in Fig.~\ref{fig5} consists of  $5$ generator nodes,  $14$ nongenerator nodes (i.e., buses), and $20$ transmission lines.
The information available on these systems includes data on the real power $P^\text{(m)}_i$  supplied by each generator, 
the real power $P^\text{(d)}_i$ requested at each node,
and the reactance $x_\ell$ for each transmission line~\cite{milano2005open}. 
In our calculations, we keep all these parameters unchanged except for the $P^\text{(m)}_i$ of the generators with zero output of real power. 
To include those generators in our formalism, we assume that they have a small real-power output 
of $1$ MW. We also assume that all  generator rotors  have identical dynamic parameters, 
with rotor inertia $M_i=5$ (in seconds) and rotor damping ratio $D_i=5$.
The transient reactance of each generator is chosen to be $0.001$ p.u.,
which guarantees that the angle difference between the two ends of the 
line connecting the generator node (and hence instability) remains small during cascades. 
Finally, the load frequency ratio $T_i$ and line capacities $W_\ell$ are assigned in the same way as in 
Iceland's power grid.

\clearpage

\bigskip
\bigskip

\begin{table}[H] 
\caption{Description of the power systems used in this work. 
The columns represent the number of  buses  ($n$),  
 generators  ($n_\text{g}$), and power lines ($n_{l}$), respectively.
} 
\centering
\begin{tabular}{lrrr}
\hline 
Power systems ~~~~~~~&~~~~~~ $n$  &~~~~~~~$n_\text{g}$ & ~~~~~~~$n_{l}$\\ 
\hline \hline
IEEE14 & $14$~ & $5$ & $20$\\ 
\hline
IEEE39 &  $39$~  & $10$ & $45$\\ 
\hline
IEEE57 &  $57$~  & $7$ & $78$\\ 
\hline
PEGASE89 &  $89$~  & $12$ & $206$\\ 
\hline
IEEE118 &  $118$~ & $54$ & $179$\\ 
\hline
Iceland &  $189$~ & $35$ & $203$\\
\hline
\end{tabular}
\label{table:results_comparison}  
\end{table}

\begin{table}[H] 
\caption{Cascades triggered by single- and double-line removal perturbations on
the test power systems considered. Here,  $N$ is the number of 
 perturbations that keep the network initially connected and $N_T$ is the number of such perturbations 
that trigger cascades. Also shown is   
the number of double-line removals $N^{\text{OM}}$ 
for which the resulting $C'$ depends on the perturbation schedule (i)-(iii)  (defined in the main text), as well as the
breakdown into the number for which the largest $C'$ results 
from either (i) or (ii)  ($N^{\text{E}}$),
from both  (i) and (ii) ($N^{\text{B}}$), 
and from (iii) ($N^{\text{C}}$).  Similar results for 
Iceland's power grid are presented in the main text.\\  
} 
\centering
\begin{tabular}{l|rc|rrrrrr}
\hline 
 \multicolumn{1}{l|}{System}  & \multicolumn{2}{c|}{Single} &   \multicolumn{6}{c}{Double}\\
 \hline \hline
  &  $N$~ &$N_T$  & ~~$N$ & ~~$N_T$ & ~$N^{\text{OM}}$& ~$N^{\text{E}}$ & ~$N^{\text{B}}$ & ~$N^{\text{C}}$\\ 
\hline \hline
IEEE14 & $19$~ & $1$ & $163$  & $45$ & $17~$ & $9$ & $8$ & $0$\\
\hline
IEEE39 &  $35$~  & $3$ & $562$  & $81$ & $43~$ & $27$ & $11$ & $5$\\
\hline
IEEE57 &  $77$~  & $3$ & $2859$ & $220$ & $83~$ & $71$ & $6$ & $6$\\
\hline
PEGASE89 &  $189$~  & $5$ & $17739$ & $1092$ & $308~$ & $245$ & $34$ & $29$\\
\hline
IEEE118 &  $170$~ & $3$ & $14289$ & $946$ & $564~$ & $455$ & $80$ & $29$\\
\hline
\end{tabular}
\label{table:table1}  
\end{table}

\begin{figure}[H]
\centering{
\includegraphics[width=6 cm]{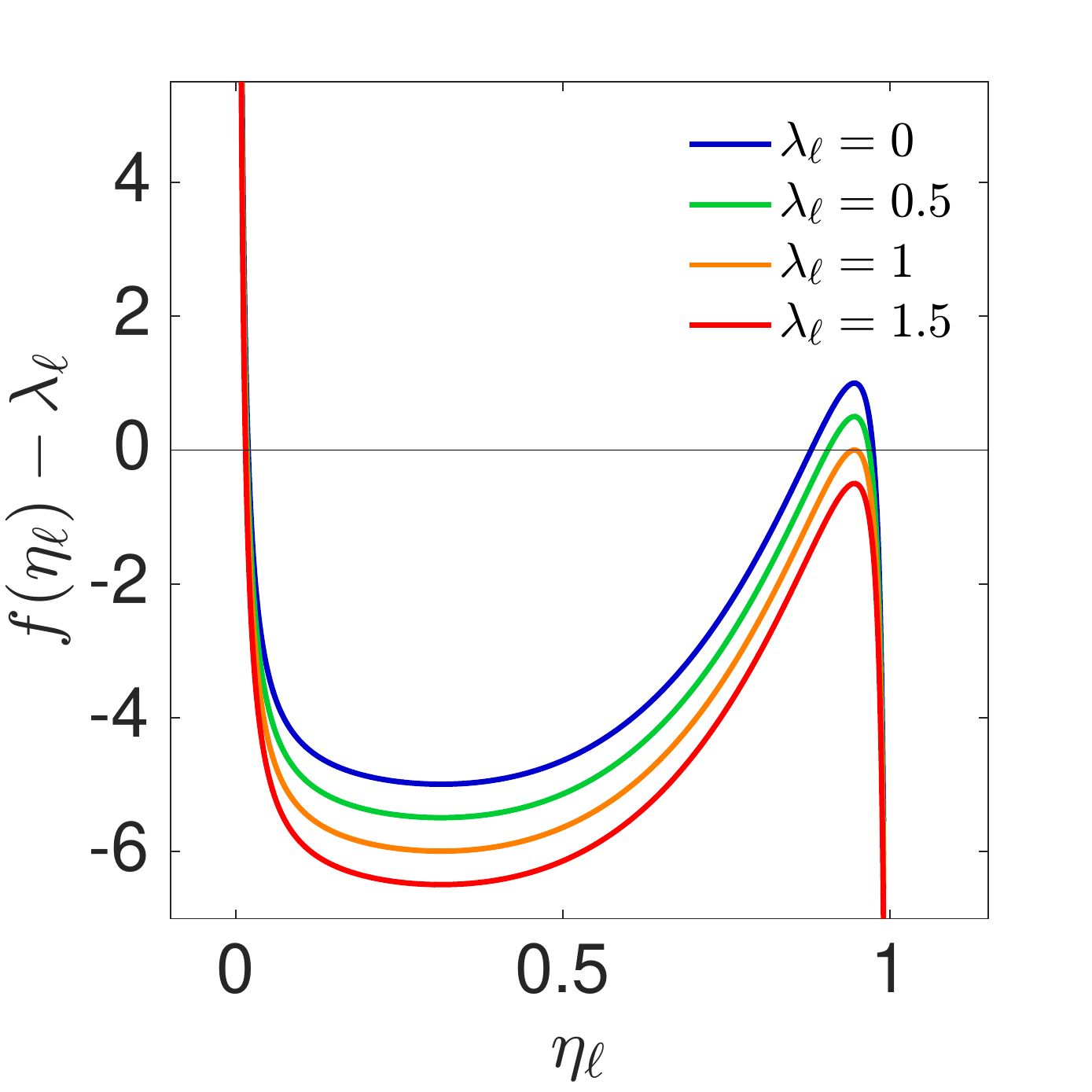}} 
\caption{ 
Equivalent of Fig.~\ref{fig1}(a) [i.e., the rhs of Eq.~\eqref{eq:eq1}] for 
different values of $\lambda_\ell$.
\label{fig_rhs}}
\end{figure}

\begin{figure}[H]
\centering{
\includegraphics[width=9 cm]{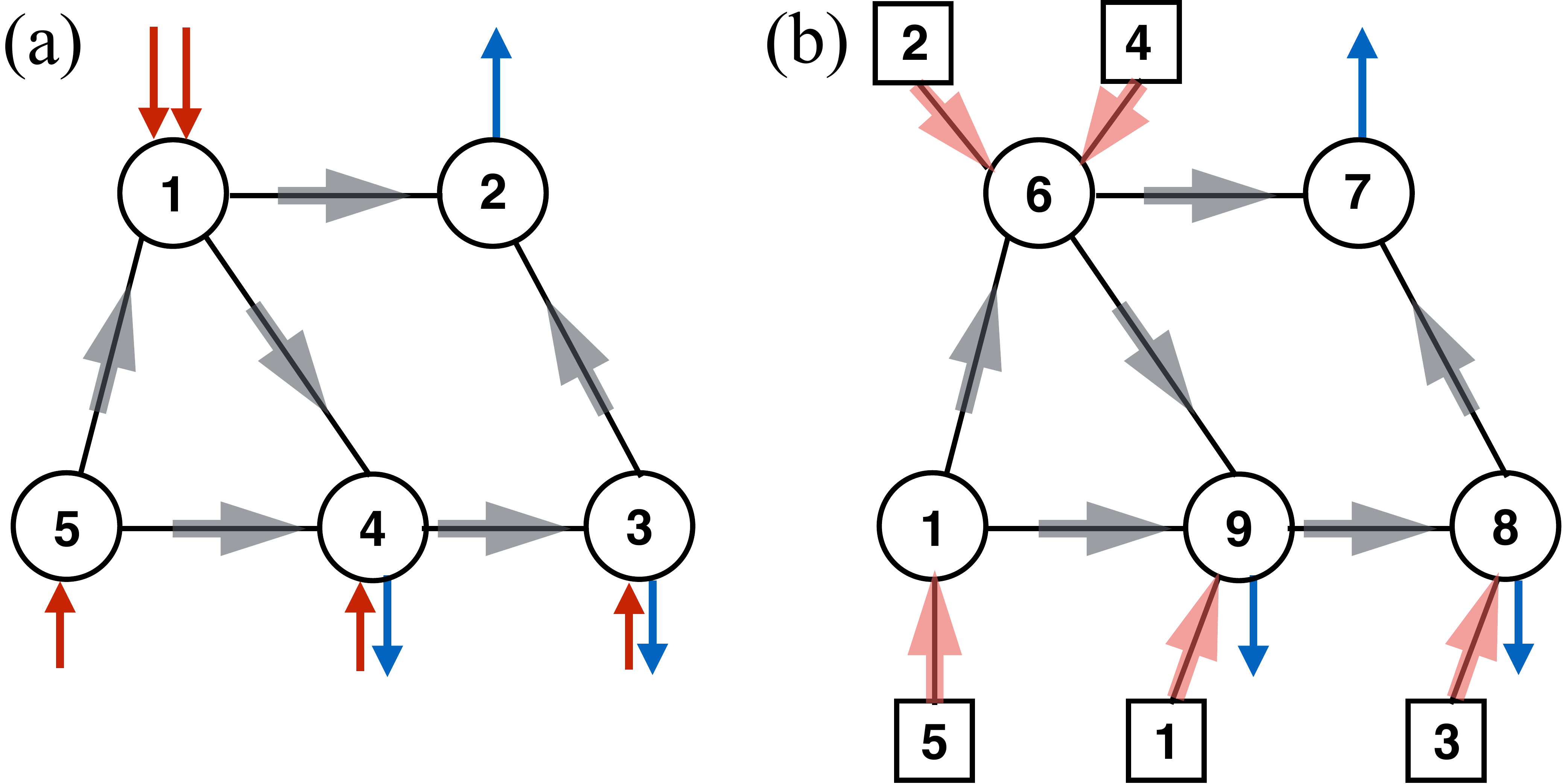}} 
\caption{Network representations for a hypothetical $5$-bus power system. (a) Original representation, where each node is associated with  a load (blue arrow) and/or a redistribution point, which may be connected to one or more generators (red arrows).
 The wide gray arrows indicate the direction of the flow on the lines. 
(b) Extended representation, where each generator is modeled as an additional node (square) that injects power into the original network through a virtual line (wide red arrow). 
\label{fig2}}
\end{figure}

\begin{figure}[ht]  
\includegraphics[width=17cm]{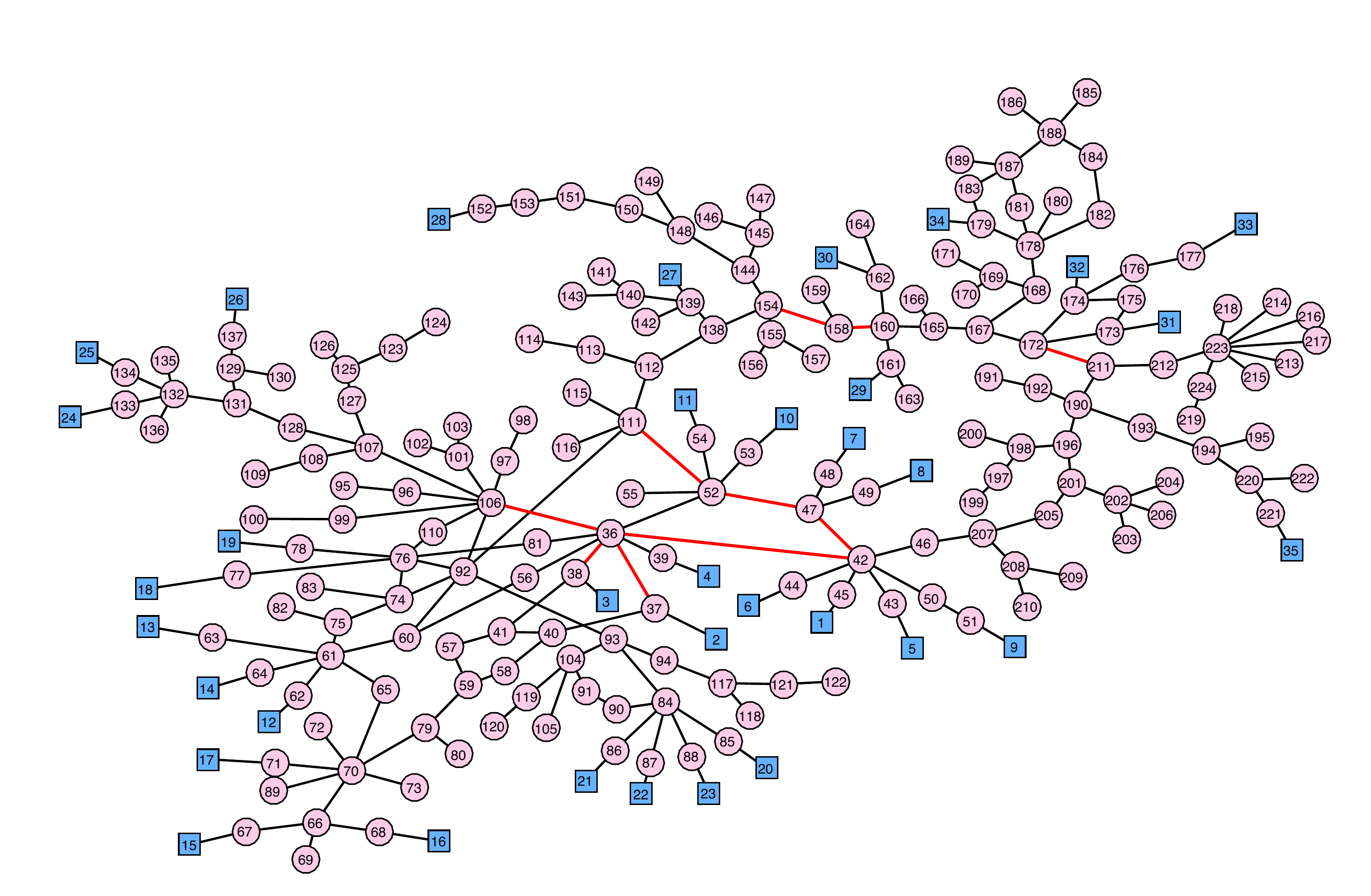}
\vspace{-0.2cm}
\caption{Network representation 
of 
Iceland's power-grid system.
 Generators are indicated by  squares and nongenerator nodes by circles.  
Under the conditions considered here, there are $10$ single-line removal perturbations (red) that lead to cascades 
in this system.  
\label{iceland_network}}
\end{figure}

\begin{figure}[ht] 
\includegraphics[width=12cm]{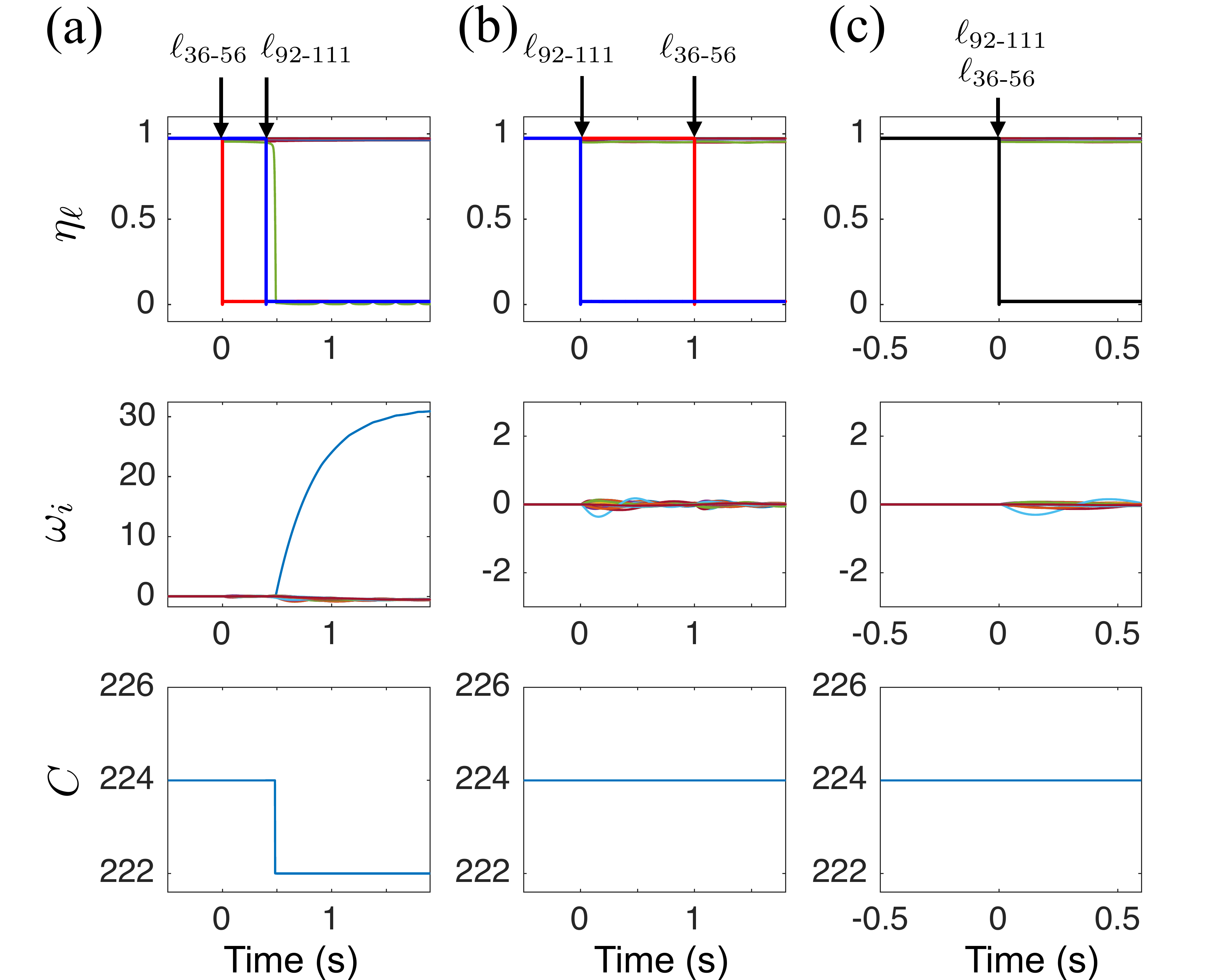}
\caption{Impact of perturbation order in 
Iceland's
power grid for the removal of lines  $\ell_{36\text{-}56}$ and $\ell_{92\text{-}111}$ according to the three scenarios considered in the main text:
(a) removal of $\ell_{36\text{-}56}$ followed by the removal of  $\ell_{92\text{-}111}$;
(b)  removal of  $\ell_{92\text{-}111}$ followed by the removal of  $\ell_{36\text{-}56}$;
(c) concurrent removal of $\ell_{36\text{-}56}$ and $\ell_{92\text{-}111}$.
The dynamics of the power grid is represented by the status of the transmission lines $\eta_\ell$ (top panels), the frequency of the generators $\omega_i$ (middle panels),  and the size of the largest cluster $C$  (bottom panel).
\label{fig4}}
\end{figure}  

 \begin{figure}[ht] 
\includegraphics[width=12cm]{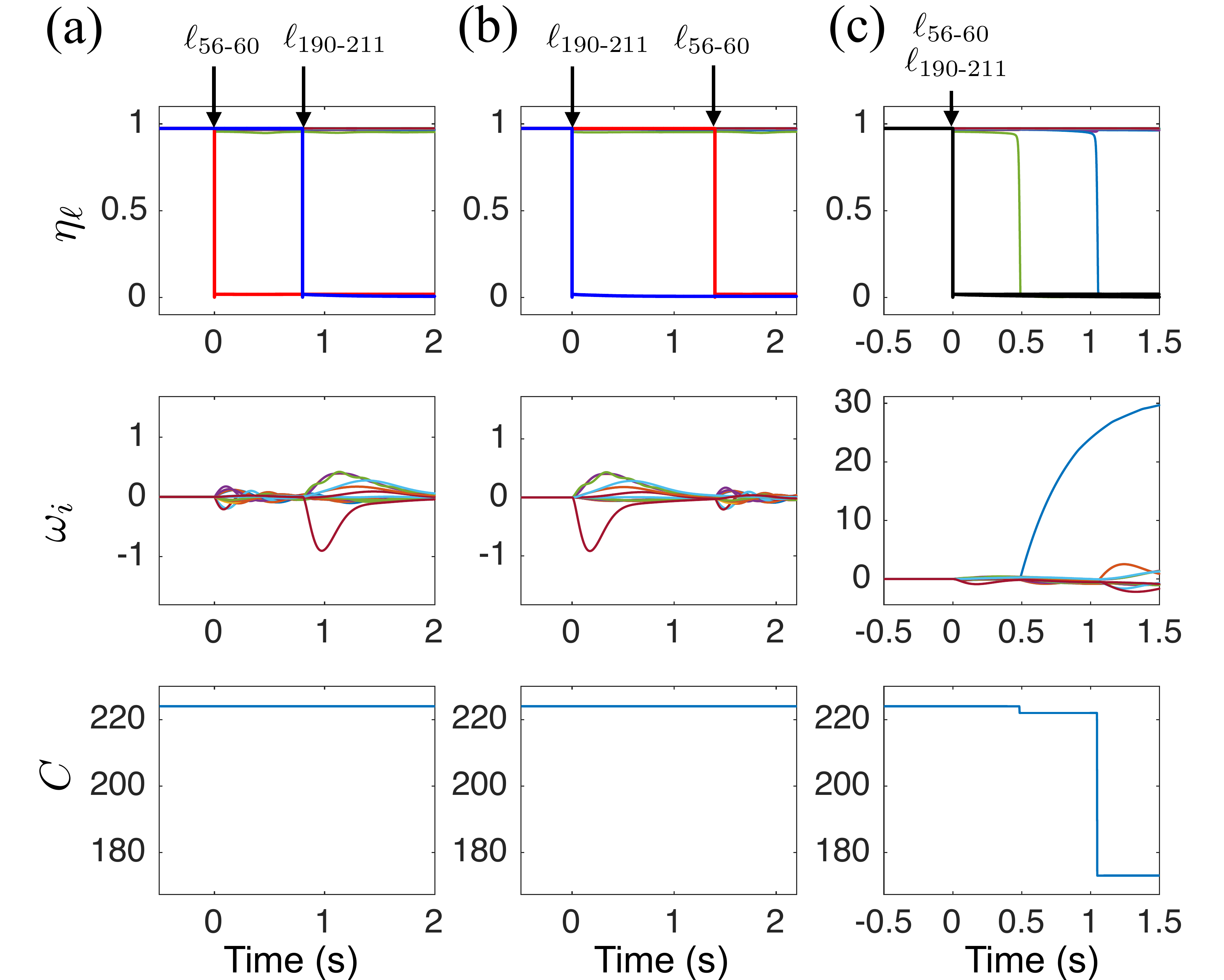}
\caption{Same as Fig.~\ref{fig4}  for the removal of lines $\ell_{56\text{-}60}$ and $\ell_{190\text{-}211}$.
In this case the concurrent removal of the two lines has the highest impact, while in the example of 
Fig.~\ref{fig4} a  cascade
can only be triggered by the separate removal of the lines.
\label{spp_orderMatter}}
\end{figure}

 \begin{figure}[h!]  
\includegraphics[width=12cm]{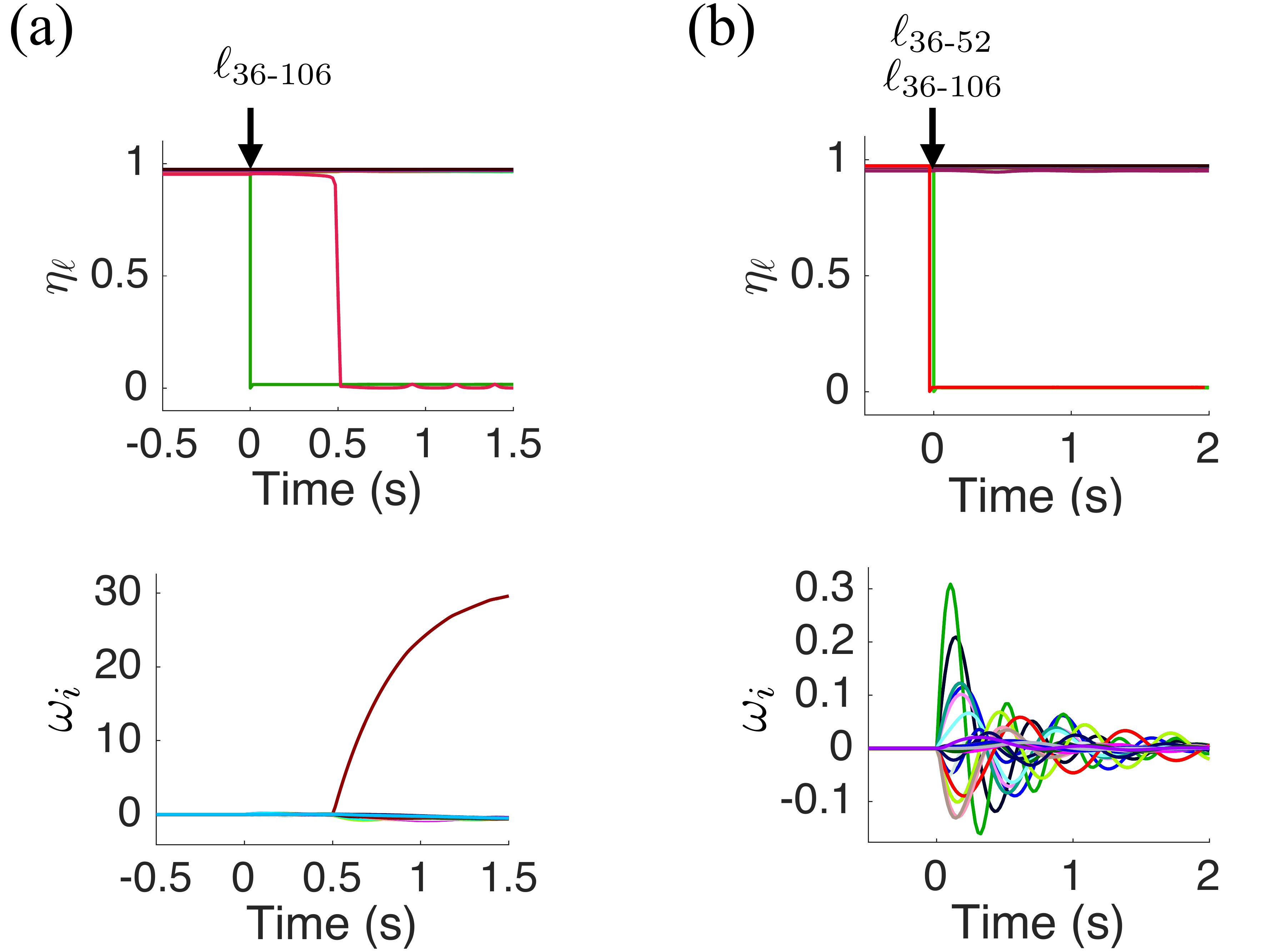}
\caption{Rescue perturbation in 
Iceland's
power-grid system.
(a) The removal of line $\ell_{36\text{-}106}$  eventually leads to the separation of generator $1$ due to the overload of line $\ell_{42\text{-}45}$,  which causes the network to lose $189\,$MW 
in power generation ($13.8\%$ of its total).
(b) The concurrent removal of line $\ell_{36\text{-}52}$ along with $\ell_{36\text{-}106}$ prevents subsequent failures and power losses, keeping the network connected.
The top panels 
show the status of the transmission lines $\eta_\ell$ and the bottom panels 
show
the frequency of the generators $\omega_i$.  
\label{spp_fig2}}
\end{figure}


\begin{thebibliography}{99}
\bibitem{data}
Theoretical understanding is especially needed given that available reports of empirical data 
on cascade blackouts  generally do not include detailed information about generator behavior along with power flow information.
\end{thebibliography}
\end{document}